# Study of Ni and Zn doped CeOFeAs: Effect on the structural transition and specific heat capacity

S. J. Singh[a], Jai Prakash[b], A. Pal[c], S. Patnaik[a], V. P. S. Awana[c] and A. K. Ganguli[b*]

[a]School of Physical sciences, Jawaharlal Nehru University, New Delhi 110067, India

[b]Department of Chemistry, Indian Institute of Technology, New Delhi 110016, India

[c]National Physical laboratory, New Delhi, Dr. K. S. Krishnan Marg, New Delhi 110012, India

We have systematically studied the substitution of nonmagnetic Zn and magnetic Ni at iron sites in Ce based oxypnictide. The parent compound (CeOFeAs) shows an anomaly in resistivity around 150 K due to structural transition from tetragonal (space group: *P*4/*nmm*) to orthorhombic structure (space group: *Cmma*). Substitution of Zn suppresses this anomaly to lower temperature (~130 K) but Ni substitution does not show any anomaly around this temperature and the compound behaves like a metal. Further, we find that non magnetic (Zn) doping leads to higher impurity scattering as compared to magnetic Ni doping. Similar to the resistivity measurement, the specific heat shows another jump near 4 K for CeOFeAs. This is attributed to the ordering of $Ce^{3+}$ moments. This peak shifts to 3.8 K for Zn substituted compound and there is no change in the ordering temperature in the Ni substituted CeOFeAs. These peaks are broadened in applied magnetic field (5 T) and the calculated magnetic entropy tends to saturate at the same value for 0 T and 5 T external magnetic field.

---

**author for correspondence: Email: ashok@chemistry.iitd.ac.in,*

*Tel : 91-11-26591511; FAX : 91-11-26854715*

## 1. Introduction

The main structural feature of the iron-arsenic based superconductor (REOFeAs; RE = rare earth) is the FeAs planes. The parent compound of these iron pnictides are antiferromagnets [1] similar to cuprate based high temperature superconductors. These compounds have a tetragonal structure and show superconductivity after suitable substitution [1-4]. Many studies have been reported on the doping at various sites of REOFeAs. In some cases the transition temperature and upper critical field increase while doping of certain ions lead to impurity scattering and pair breaking that do not support the superconducting phases.

The emphasis of current discussion in these new superconductors is the origin of the superconducting pairing mechanism. Theoretical studies suggest that the pairing may be realized via inter-pocket scattering of electrons between the hole and electrons pockets, giving rise to a multiband s-wave pairing with possible sign change of the superconducting gap on the Fermi surface (so called s± gap) [5-7]. Some other models have suggested that the pairing mechanism has its origin in magnetic super exchange [8-9] with multiband s-wave pairing. Further, it has also been established that if the height of pnictogen from Fe-plane is varied, the pairing symmetry could effectively be changed from s-wave to d-wave symmetry [10]. The experimental results are equally controversial. While some experiments support the s± symmetry, some other reflects d-wave symmetry [11-17]. These controversies have necessitated the study of impurity scattering and pair breaking through magnetic and non-magnetic doping in the oxypnictides in greater detail.

In this context, Anderson model suggests that while in conventional s-wave superconductors the non-magnetic dopant would not lead to pair breaking [18], in d-wave superconductors (as seen in cuprates), the substitution of such impurities could lead to rapid suppression of superconductivity [19]. For example, it is now established that doping of non-magnetic $Zn^{2+}$ in cuprates, suppresses the transition temperature more effectively in comparison to magnetic impurities like Mn or Co [20]. With regard to ferropnictides, band structure calculations [21] suggest that the iron based superconductors have itinerant behaviour of Fe 3d electrons. In contrast, the copper (Cu) 3d electrons have localized behaviour in cuprate superconductors. Based on these ideas, we have reported the substitution of cobalt at iron sites in CeOFeAs to introduce extra electrons and induce superconductivity in Co-doped CeOFeAs [22]. Superconductivity is also observed in nickel doped LaOFeAs [23]. Li *et al* [24] have shown that the substitution of nonmagnetic Zn ions (up to ~10%) at iron sites does not affect the transition temperature of 10% F-doped

LaOFeAs. On the contrary similar study by Guo et al. has indicated large decrease in $T_c$ of LaOFeAs with 3% Zn doping [25]. Further, no such study on Ce based ferropnictides has been reported. This is of interest because unlike the LaOFeAs, the parent Ce-based oxypnictide shows two antiferromagnetic transitions one around 150 K due to Fe 3d electrons, and the other around 4 K due to ordering of $Ce^{+3}$ ions. In this paper, we have studied the effect of partial substitution of the nonmagnetic Zn and magnetic Ni ions at iron sites in CeOFeAs and have investigated its structural, electronic and magnetic properties. Our result suggests that the substitution of non-magnetic ions generate higher disorder induced impurity scattering as compared to magnetic ion doping.

## 2. Experiments

For the synthesis of $CeOFe_{1-x}M_xAs$ (M = Ni and Zn), high purity Ce, As, $CeO_2$, Ni, ZnO, and FeAs were used. FeAs was obtained by heating stochiometric amounts of Fe chips and As powder in an evacuated silica tube at 595°C for 12 hours followed by heat treatment at 630ºC for 12 h and finally heated at 800-900ºC for 24 hours. The rare earth oxides were preheated at 900°C before weighing. The reactants were weighed according to the stoichiometric ratio in a $N_2$-filled glove- box and then sealed in evacuated silica ampoules ($10^{-4}$ torr) and heated at 950°C for 48 hours at a rate of 50°C/h. The resulting powder was compacted into disks under 5 ton pressure. The disks were wrapped in Ta foil, sealed in evacuated silica ampoules and annealed at 1150°C for 48 hours at a rate of 100°C/h and then cooled to room temperature. Powder X-ray diffraction patterns of the finely ground powders were recorded with Cu-Kα radiation in the 2θ range of 20° to 70°. The lattice parameters were obtained from a least squares fit to the observed *d* values.

Resistivity measurement was carried out using a Cryogenic 8 T Cryogen-free magnet in conjunction with a variable temperature insert (VTI). The samples were cooled in helium vapour and the temperature was measured with an accuracy of 0.05 K using a calibrated Cernox sensor wired to a Lakeshore 340 temperature controller. Standard four probe technique was used for transport measurements. The external magnetic field (0-5 T) was applied perpendicular to the probe current direction and the data were recorded during the warming cycle with heating rate of 1 K/min. The magnetic and specific heat measurements were carried out on Quantum design Physical property Measurement System (PPMS).

## 3. Results and Discussion

Powder X-ray diffraction patterns for Ni substituted $CeOFe_{1-y}Ni_yAs$ (‘y’ = 0, 0.1 and 0.2) are shown in figure 1(a). All observed reflections could be satisfactorily indexed on the basis of tetragonal crystal structure (space group: *P*4/*nmm*). The variation of lattice parameters (**a** and **c**) with nickel substitution is shown in figure 1(b). The **c**-lattice parameter decreases with increase in nickel substitution which is expected since the ionic size of $Ni^{2+}$ (0.55 Å) is smaller as compared to $Fe^{2+}$ (0.63 Å) in tetrahedral coordination. The variation of **a**-lattice parameter with Ni doping is not systematic. For 'y' = 0.1 composition, a slight decrease in **a**-lattice parameter is observed while the 'y' = 0.2 composition has approximately the same **a**-lattice parameter as for 'y' = 0 composition. The c/a ratio and volume of tetragonal cell also shrinks on Ni doping in $CeOFe_{1-y}Ni_yAs$. Cao *et al* [23] also reported reduction in **c**- lattice parameter, c/a ratio and volume of cell on Ni substitution in LaOFeAs.

Powder X-ray diffraction patterns of Zn doped $CeOFe_{1-x}Zn_xAs$ ('x' = 0, 0.15, 0.2) are shown in figure 1(c). For 'x' = 0.15 compositions, all the observed reflections could be satisfactorily indexed on the basis of the tetragonal CeOFeAs cell (space group: *P*4/*nmm*). The 'x'= 0.2 composition shows presence of small amount of $CeO_2$ (~5%) along with the major tetragonal superconducting phase. The **a** and **c**-lattice parameters decrease with zinc substitution in $CeOFe_{1-x}Zn_xAs$ as shown in figure 1(d) which could be attributed to smaller ionic size of $Zn^{2+}$ ion (0.60 Å) as compared to $Fe^{2+}$ ion (0.63 Å). The lattice parameters for Zn substituted $CeOFe_{1-x}Zn_xAs$ are slightly larger as compared to the Ni substituted analogues which is expected since $Zn^{2+}$ is larger as compared to $Ni^{2+}$.

Figure 2 shows the temperature dependence of resistivity for Zn and Ni doped CeOFeAs. The undoped parent compound (CeOFeAs) shows a resistivity anomaly below 170 K as shown in figure 2(a) which is related to structural transition followed by magnetic ordering. On addition of 15% and 20% of Zn, this anomaly is shifted to lower temperatures of 135 K and 127 K respectively, as shown in the inset of figure 2(a). Below this temperature the resistivity behaviour is like that of a semiconductor. The room temperature value of the resistivity was found to be 6.5 mΩ-cm, 13.5 mΩ-cm and 13.1 mΩ-cm for CeOFeAs, $CeOFe_{0.85}Zn_{0.15}As$ and $CeOFe_{0.8}Zn_{0.2}As$ respectively. Above the anomaly temperature, the resistivity is approximately same for both the Zn-doped samples. However, below the anomaly the resistivity increases rapidly for the 20% Zn doped sample which clearly suggests the enhanced disorder due to Zn doping.

The Ni-doped CeOFeAs compounds (10% and 20% Ni) behave like metals having a linear temperature dependence of the resistivity but do not show any anomaly around 150 K like the parent compound. Both samples show a drop in resistivity around 4.5 K and 3.4 K for 10% and 20% Ni-doped samples respectively but do not show zero resistivity till 1.6 K. This drop in resistivity may also be due to the antiferromagnetic ordering of $Ce^{+3}$ similar to that known for the parent compound (inset of figure 2(b)). Cao *et al* [23] have reported that Ni doping in LaOFeAs shows superconductivity below 5% Ni doping. At higher Ni-doping it shows the drop in resistivity similar to seen in our studies. One main inference is that doping of Zn in place of Fe, increases the resistivity of parent compound whereas Ni doping makes it more metallic. To study the state of the magnetic transition with Zn and Ni doping, magnetization as a function of temperature under field cooled and zero field cooled protocol was carried out (not shown). It was observed that the magnetic transition appeared at lower temperature as compared to undoped CeOFeAs and the suppression was much more pronounced in Ni doped sample.

In the following, we address these results in the light of band-structure calculations in similar oxypnictides. The energy dependence of the calculated electron density of states (DOS) for LaO(Fe/M)As (M= Mn, Co and Ni) [26] show that partial substitution of $Fe^{2+}$ ($3d^6$) by $Ni^{2+}$ ($3d^8$) does not change the total DOS but increases the chemical potential. As a result the top of valence band are pushed down to the Fermi level with band filling (adding electrons) indicating that 3d electrons of both Fe and Ni ions in this ZrCuSiAs-type structure have the itinerant character. Also LaONiAs [27-28] is more metallic and has a higher charge carrier density compared to its iron counterpart. Thus a more metallic state in $CeOFe_{1-x}Ni_xAs$ is expected with increasing x because Ni doping leads to direct injection of carriers in FeAs layer itself and is equivalent to electron doping, leading to a drop in resistivity with Ni doping as shown in figure 2(b). Cao *et al*. [23] and Li *et al*. [29] have also reported the electronic phase diagram for $LaOFe_{1-x}Ni_xAs$ and $SmOFe_{1-x}Ni_xAs$ respectively, and shows that resistivity decreased with Ni doping and no anomaly was observed for $x>0.05$. Their results are consistent with our studies on Ni doping in CeOFeAs. Unlike the magnetic $Ni^{2+}$ ion, $Zn^{2+}$ ion ($3d^{10}$) is non-magnetic and hence Zn doped CeOFeAs compounds exhibit different physical properties as compared to Ni doped compounds. Band structure calculation of LaOZnAs [30] depicts that the fully occupied Zn 3d states are located about -7 eV to the Fermi level and are separated from the near-Fermi valence band by a gap i.e. the 3d electrons of Zn in this structure are localized. Thus the partial substitution Zn at Fe site is not suppose to add more itinerant electrons into carrier conducting FeAs layers. Its main contribution would be to cause disorder which is the reason for such a dramatic difference in physical properties of Zn doped compounds compared to their Ni doped analogues. This feature is quite different from high $T_c$ cuprates, where even small levels of chemical substitution of $Zn^{+2}$ into the $CuO_2$ planes results in suppression of superconductivity [31].

The temperature dependence of specific heat for all Ce-based samples is shown in figure 3. There is a clear specific heat jump close to 150 K for the x = 0 sample (CeOFeAs) as shown in inset of figure 3(a). A similar jump in the specific heat has been reported in LaOFeAs [32] and SmOFeAs [33]. The specific heat jump around 150 K is consistent with the resistivity anomaly as shown in figure 2. Similar to La and Sm based samples, this specific heat jump and resistivity drop (in the Ce- analog) around 150 K is related to the structural transition. No such demarcation is clearly indicated in Zn and Ni doped samples possibly due to paucity of data points in the relevant temperature range. Figure 3(a) also shows the specific heat data of CeOFeAs in 5 Tesla field. There is no shift in the temperature of the specific heat anomaly when compared to the data at zero magnetic field.

In the following, we focus on the low temperature behaviour of specific heat. We have measured the specific heat data for three samples, namely, the parent compound (CeOFeAs), 20% Zn-substituted phase and 10% Ni substituted phase at zero field and at 5 Tesla. In inset of figure 3(a), the specific heat for parent compound shows a sharp peak at 4.1 K. With electron doping by nonmagnetic Zn, the peak shifts to 3.8 K and its height decreases as shown in inset of figure 3(b). But substitution of Ni at Fe site does not shift the specific heat peak (4.1 K) similar to parent compound but peak height increases as shown in the inset of figure 3(c). These results support the resistivity data that the substitution of non magnetic (Zn) ions at Fe sites creates relatively higher disorder in comparison of magnetic element Ni substitution. This low temperature peak has not been seen in previous specific heat studies of La-based oxypnictides [32]. Since the only difference between these two materials is the rare earth ion (nonmagnetic $La^{+3}$ and magnetic $Ce^{+3}$) these peaks appear to be related to the magnetic ordering of $Ce^{+3}$ ions. In fact similar specific heat behaviour at low temperatures has been observed in Sm-based oxypnictide [33].

Figure 4 confirms that the low temperature anomaly is sensitive to the external magnetic field. The peak for the parent compound shifts from 4.1 to 3.7 K with applied magnetic field and the anomaly becomes broader as reported by Riggs *et al* [34]. Similar results are observed for Zn and Ni doped samples as shown in inset of figure 3(b)-(c). A comparison of the properties of various compositions is shown in figure 4.

Due to the specific heat peak at low temperature, it is not possible to extrapolate the specific heat data and obtain the coefficient of the electronic specific heat ($\gamma$). We have studied the C/T vs $T^2$ plot for CeOFeAs, Zn and Ni-doped CeOFeAs, and the calculated parameters are listed in Table I. The data of Ni- doped CeOFeAs (figure 5) between 10 and 20 K can be fit to $C/T = \gamma + \beta T^2$ which gives a value of $\gamma$ = 101 and 108 mJ/mol $K^2$ for H =

0 and 5 T respectively, as listed in Table I. This $\gamma$ value is much higher than that reported for La based pnictide ($T_c$ = 28 K and $\gamma$ = 1 mJ/mol-$K^2$ [35] but closed to Sm based superconductor ($T_c$ = 54 K and $\gamma$ = 81 mJ/mol-$K^2$) [36]. Since we have used data at relatively higher temperature and over a small temperature range (10-20 K), hence this large value of $\gamma$ may not be highly accurate. However, it gives us an idea of the coefficient of electronic heat capacity. Similarly we have obtained the values of $\gamma$ for CeOFeAs which are 199 mJ/mol-$K^2$ (H = 0) and 203 mJ/mol-$K^2$ (H = 5 T) whereas for Zn-doped CeOFeAs the values are 149 mJ/mol-$K^2$ (H = 0 T) and 158 mJ/mol-$K^2$ for (H = 5 T) respectively. The respective $\beta$ value for all samples is also listed in Table I. Using the obtained value of $\beta$ and the relation $\Theta_D = (234zR/\beta)^{1/3}$ [37], where z is the number of atoms per formula unit and R is the gas constant, we obtain the Debye temperature ($\Theta_D$) for present samples (table 1). Further, by subtracting the non-magnetic specific heat ($C_{magnetic}$) for all three compositions the magnetic contribution to specific heat can be derived;

$$C_{nonmagnetic} = \gamma T + \beta T^3$$

$$C_{magnetic} = C - C_{nonmagnetic}$$

For calculation of $C_{nonmagnetic}$, we have used $\gamma$ = 199 mJ/mol-$K^2$ and $\beta$ = 0.13 mJ/mol-$K^4$ for CeOFeAs, $\gamma$ = 149 mJ/mol-$K^2$ and $\beta$ = 0.15 mJ/mol-$K^4$ for Zn-doped and $\gamma$ = 101 mJ/mol-$K^2$ and $\beta$ = 0.15 mJ/mol-$K^4$ for Ni-doped sample obtained from fitting as shown in figure 5.

The magnetic entropy for this transition can be calculated by

$$S_{magnetic} = \int \frac{C_{magnetic}}{T} .dT$$

As the temperature increases the magnetic entropy also increases (figure 6) and then saturates above the magnetic transition. It is clear that this saturation value is approximately equal to 0.5R (R = gas constant) for all three compositions (inset of figure 6) as expected for the doublet ground state of $Ce^{+3}$ [34] which is lower than the value of Rln2 for $Sm^{+3}$ [38-39]. This suggests that the peak near 4 K can be related to the antiferromagnetic ordering of $Ce^{+3}$ ions as reported for other magnetic rare earth systems [38-39] and this peak is sensitive with respect to substitution at iron sites and also with the magnetic field. The relatively high value of '$\gamma$' may be due to interaction of the electronic wave function with $Ce^{+3}$ ions as reported elsewhere for Sm-based ferropnictides [33]. This is also reflected in the resistivity measurement which gives a hump near the antiferromagnetic transition temperature of $Ce^{+3}$.

## 4. Conclusions

Our resistivity studies show that doping of nonmagnetic Zn ions suppressed the SDW transition to lower temperature (~130 K) whereas Ni substituted CeOFeAs shows metallic behaviour (with no anomaly). Zn-substituted compositions show the semi-metallic behaviour below the anomaly (~130 K) which is due to disorder created in the FeAs layer by the substitution of Zn. The specific heat of CeOFeAs also shows a jump (sudden increase) near 4 K due to the ordered magnetic moment of $Ce^{+3}$ similar to resistivity measurements and the temperature of this jump shifted to 3.6 K on Zn doping but remains the same for Ni substituted samples. Our results suggest that this peak is sensitive with respect to substitution at iron site and external magnetic field, and the width of the specific heat peak gets broadened due to the magnetic field. The electronic contribution to the specific heat and the magnetic contribution to the entropy corresponding to the low temperature ordering of $Ce^{+3}$ have been estimated.

## Acknowledgements

AKG and SP thank DST, Govt. of India for financial support. JP and SJS thank CSIR, Govt. of India, for fellowships.

## Figure Captions:

**Figure 1.** **(a)** Powder X-ray diffraction (P-XRD) patterns and **(b)** variation of lattice parameters (‘**a**’ and ‘**c**’) for $CeOFe_{1-y}Ni_yAs$ (y = 0, 0.1, 0.2).
**(c)** PXRD patterns and **(d)** variation of lattice parameters (‘**a**’ and ‘**c**’) for $CeOFe_{1-x}Zn_xAs$ (x = 0.15, 0.2).
**Figure 2.** **(a)** Temperature dependence of resistivity for $CeOFe_{1-x}Zn_xAs$ (x = 0, 0.15 and 0.2) up to 250 K. Black arrow shows the anomaly temperature ($T_{an}$) for CeOFeAs. Inset shows the variation of resistivity anomaly temperature ($T_{an}$) with Zn content (x).
**(b)** The temperature dependence of resistivity up to 250 K for $CeOFe_{1-y}Ni_yAs$ (y = 0.1 and 0.2). Inset shows the resistivity for CeOFeAs below 7 K.
**Figure 3.** The specific heat variation with respect to temperature for **(a)** CeOFeAs **(b)** $CeOFe_{0.8}Zn_{0.2}As$ **(c)** $CeOFe_{0.9}Ni_{0.1}As$. Inset shows the behaviour of specific heat in the range of 2-5 K.
**Figure 4.** The variation of specific heat of Zn and Ni doped CeOFeAs with respect to parent compound below 10 K in presence of 0 T (closed symbol) and 5 T (open symbol) respectively**.**
**Figure 5.** Plot of C/T vs $T^2$ for $CeOFe_{0.9}Ni_{0.1}As$ with linear fit between 10 to 20 K.
**Figure 6.** Temperature dependence of magnetic specific heat ($C_m$) for CeOFeAs, $CeOFe_{0.8}Zn_{0.2}As$ and $CeOFe_{0.9}Ni_{0.1}As$ at H = 0 T. The inset shows the entropy (S) associated with the magnetic transition at H = 0 T and 5 T.

**Table 1**. Fitting parameters from the specific heat data of CeOFeAs, $CeOFe_{0.9}$ $Ni_{0.1}As$ and $CeOFe_{0.8}Zn_{0.2}As$ under 0 T and 5 T.

| **Sample** | **γ** (mJ/mol $K^2$) | **β** (mJ/mol $K^4$) | **$\Theta_D$** (K) | **S** (J/mol K) |
|---|---|---|---|---|
| CeOFeAs (0T) | 199 (±28) | 0.13 (±0.01) | 387 | 3.329 |
| CeOFeAs (5T) | 203(±28) | 0.13 (±0.01) | 388 | 2.881 |
| $CeOFe_{0.90}Ni_{0.10}As$ (0T) | 101(±23) | 0.15(±0.08) | 375 | 4.861 |
| $CeOFe_{0.90}Ni_{0.10}As$ (5T) | 108(±11) | 0.17(±0.03) | 358 | 3.636 |
| $CeOFe_{0.80}Zn_{0.20}As$ (0T) | 149(±13) | 0.15(±0.05) | 370 | 2.899 |
| $CeOFe_{0.80}Zn_{0.20}As$ (5T) | 158(±10) | 0.16(±0.03) | 368 | 2.633 |

**Figure 1**

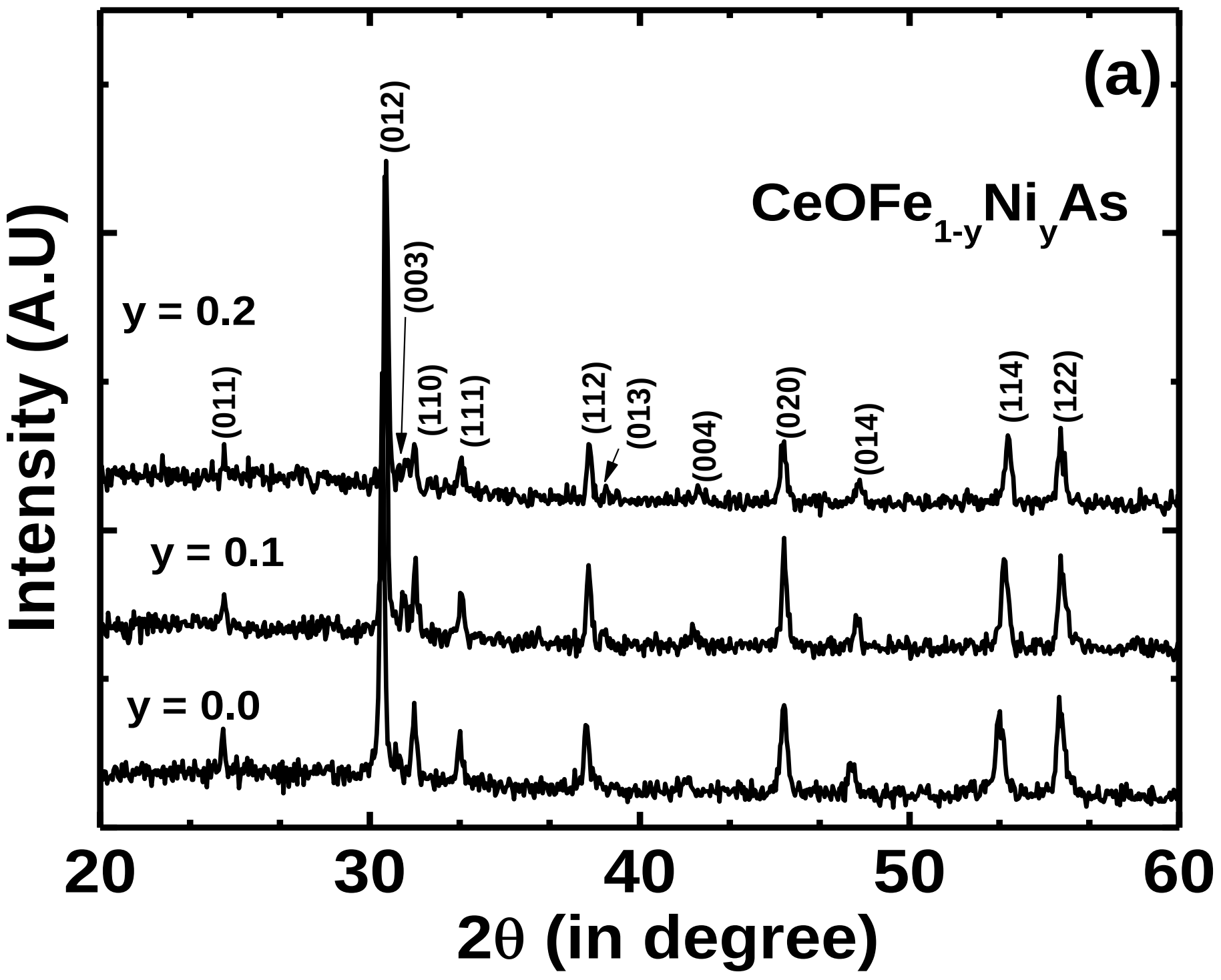

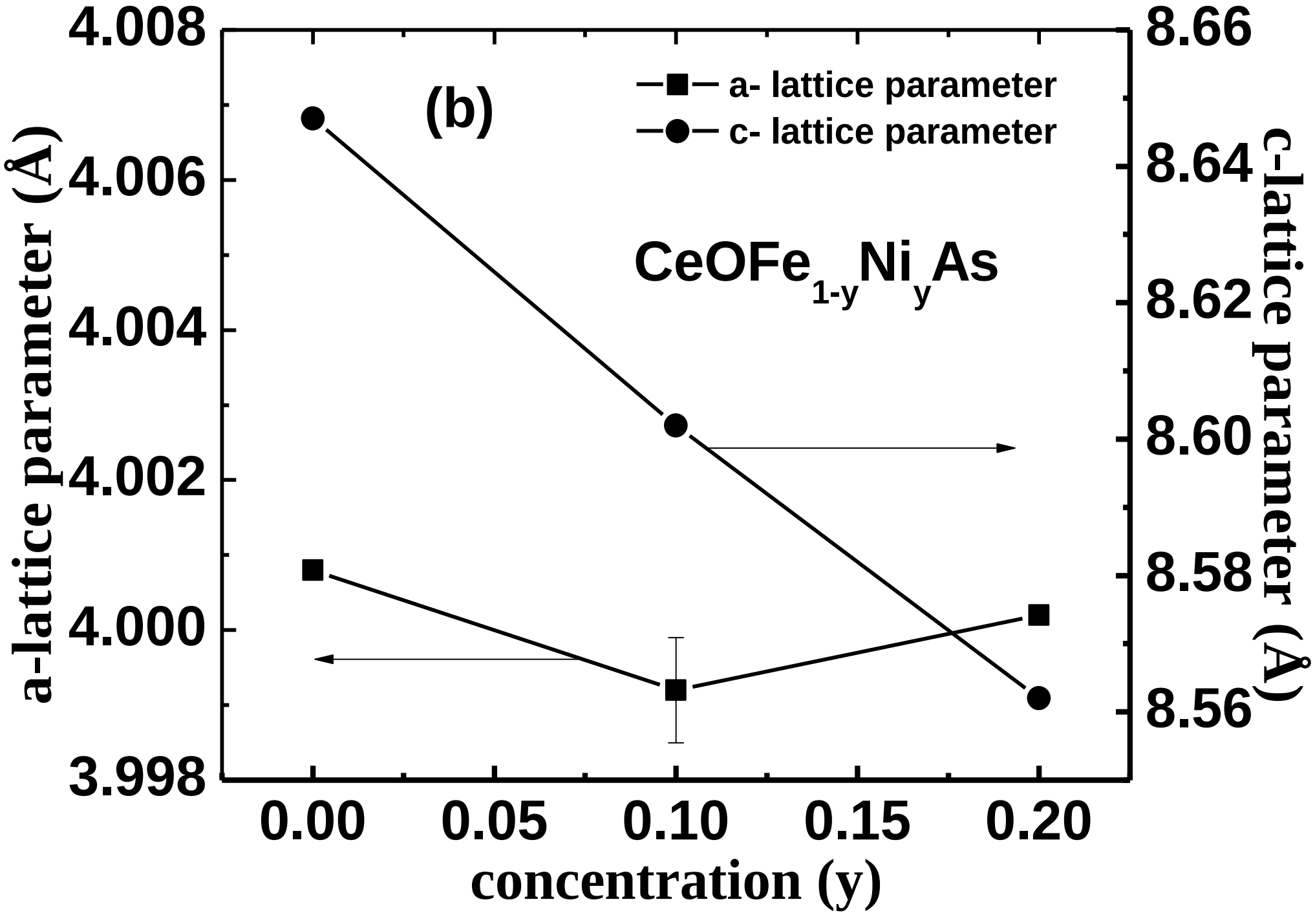

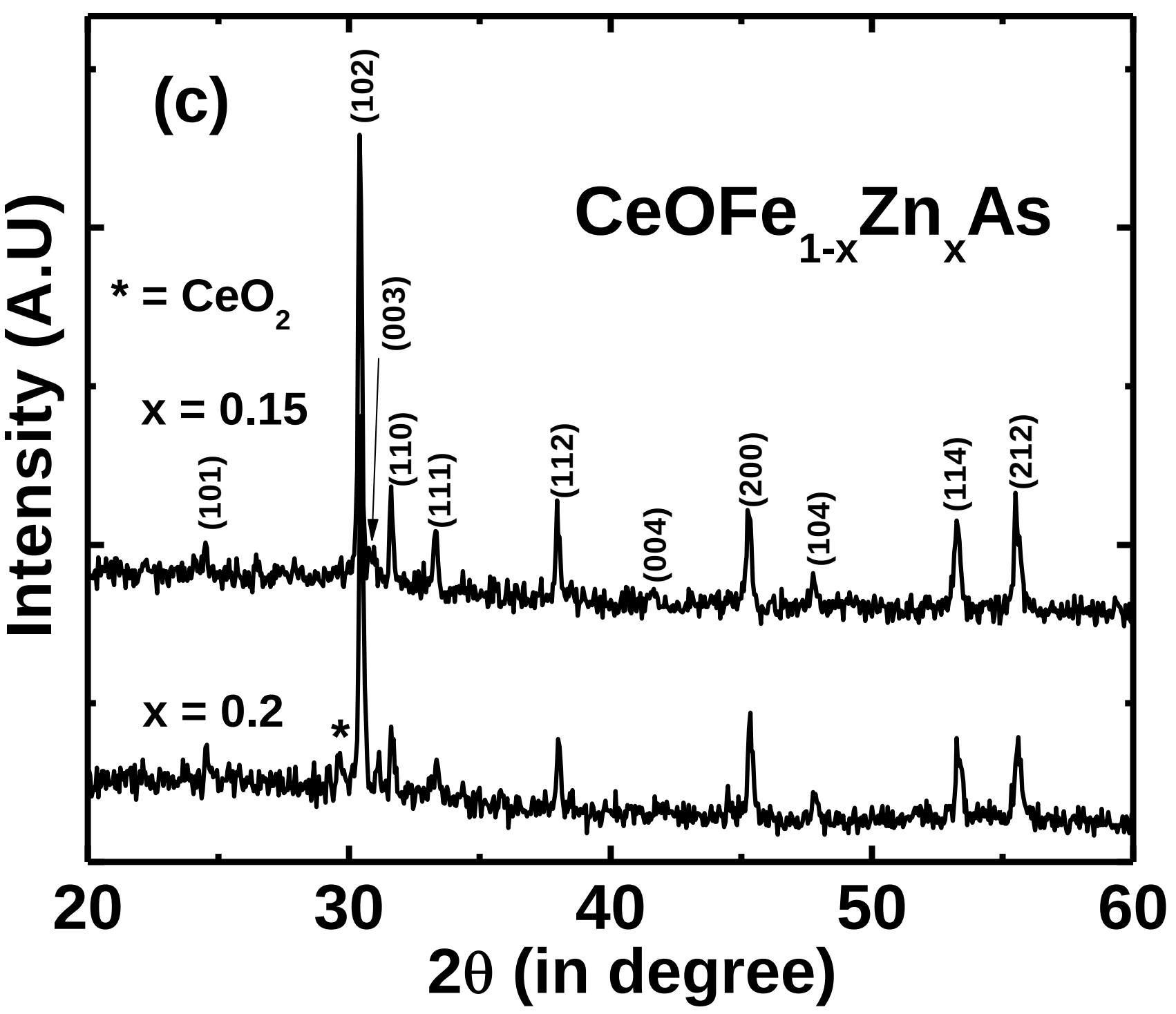

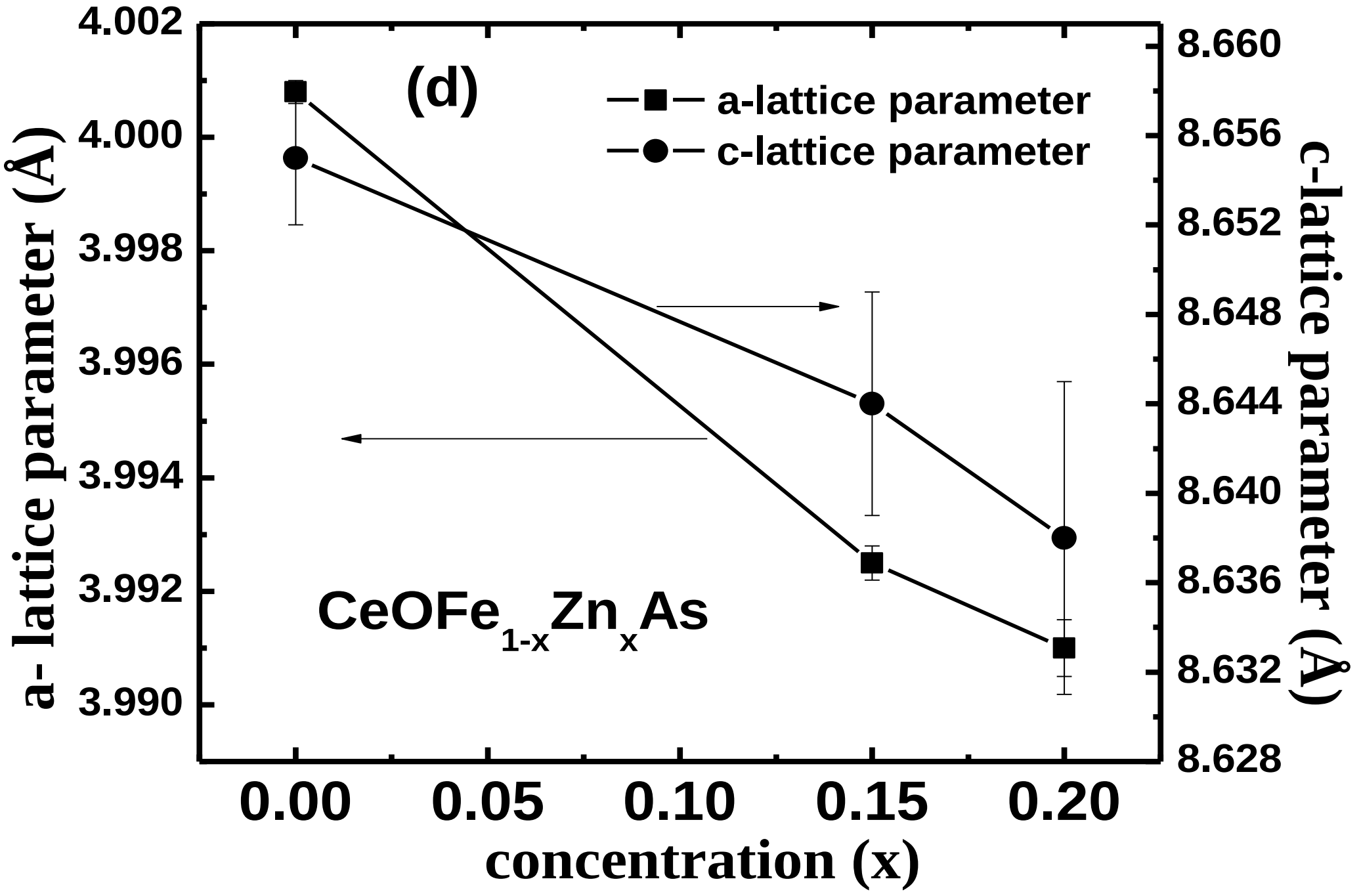

**Figure 2**

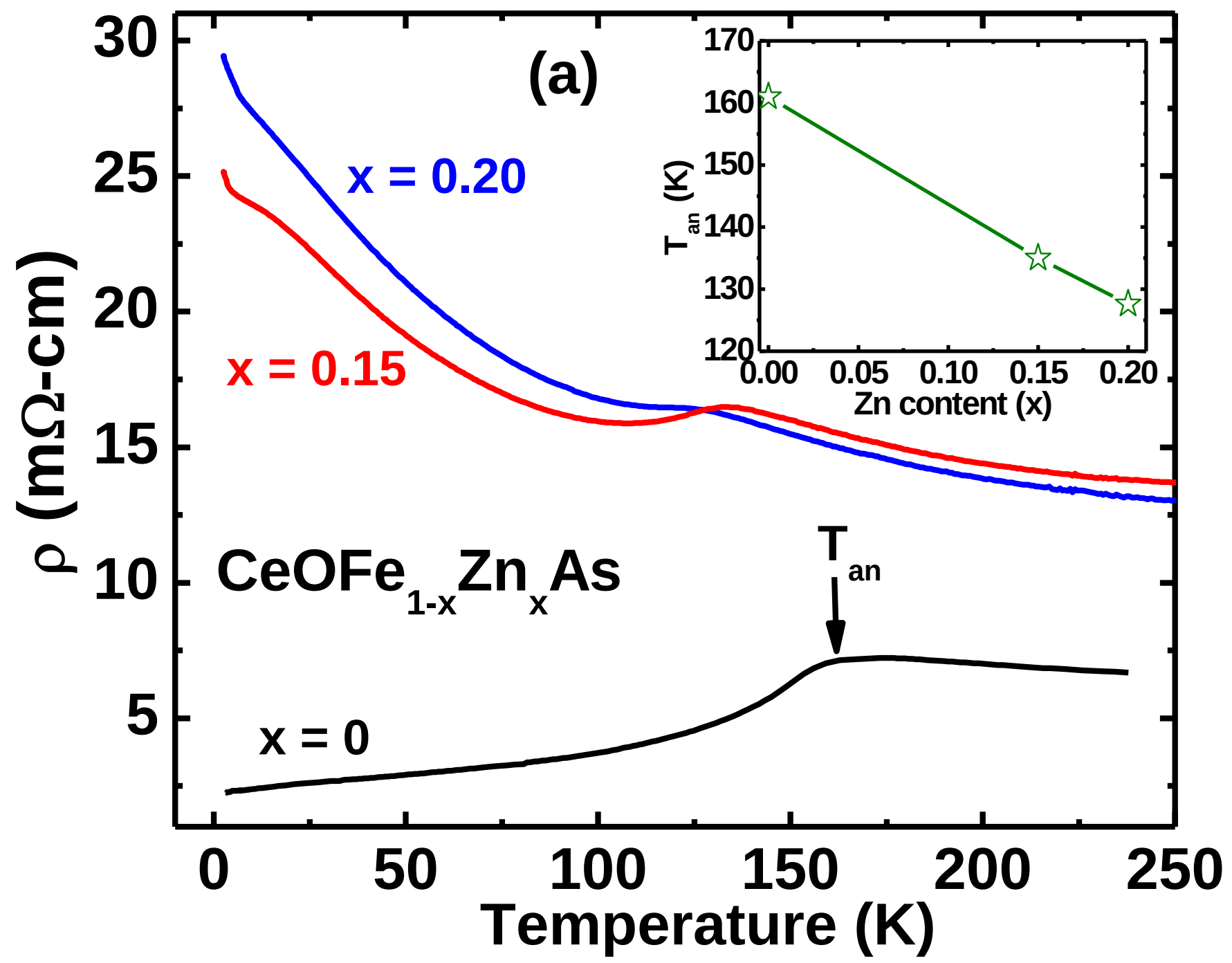

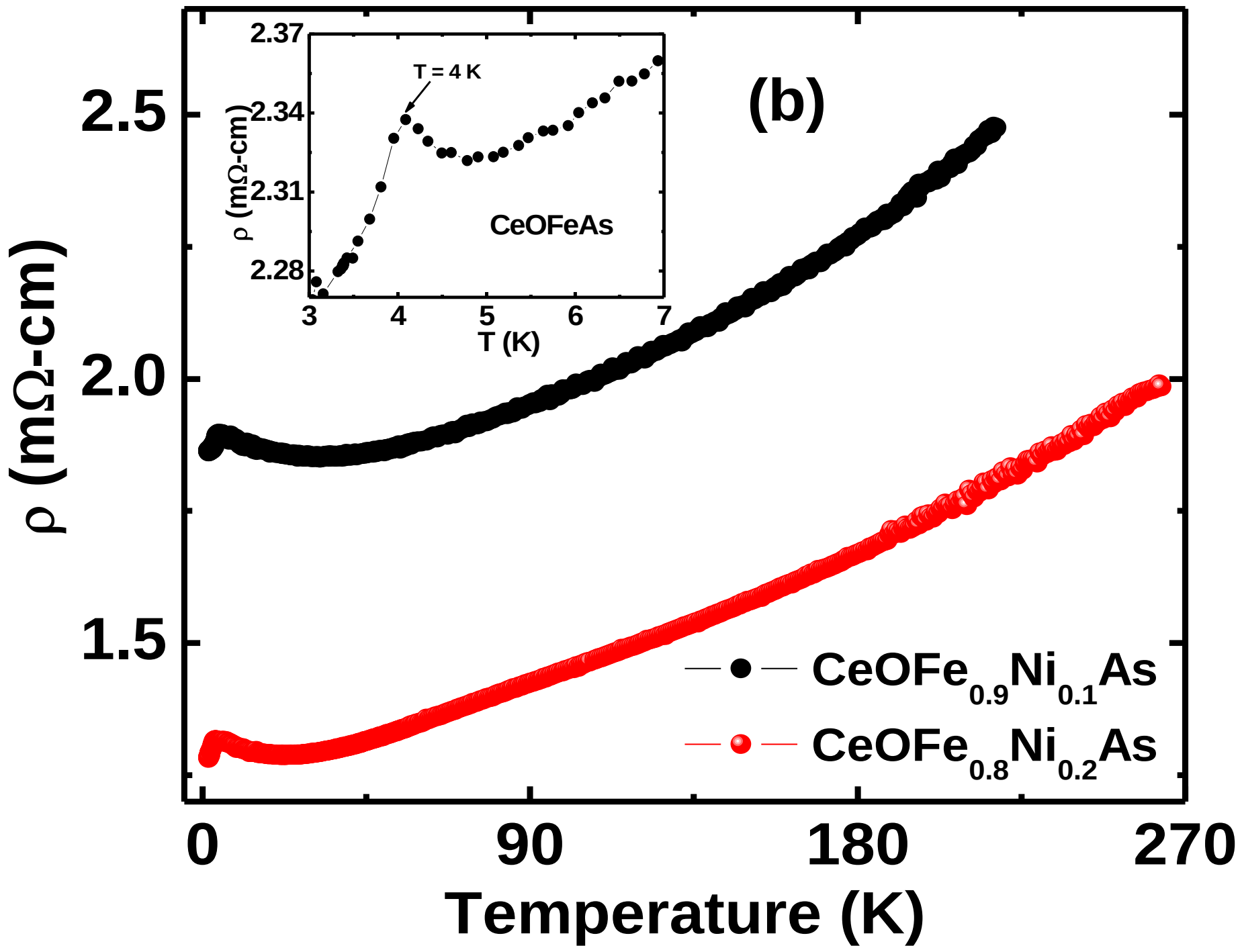

**Figure 3**

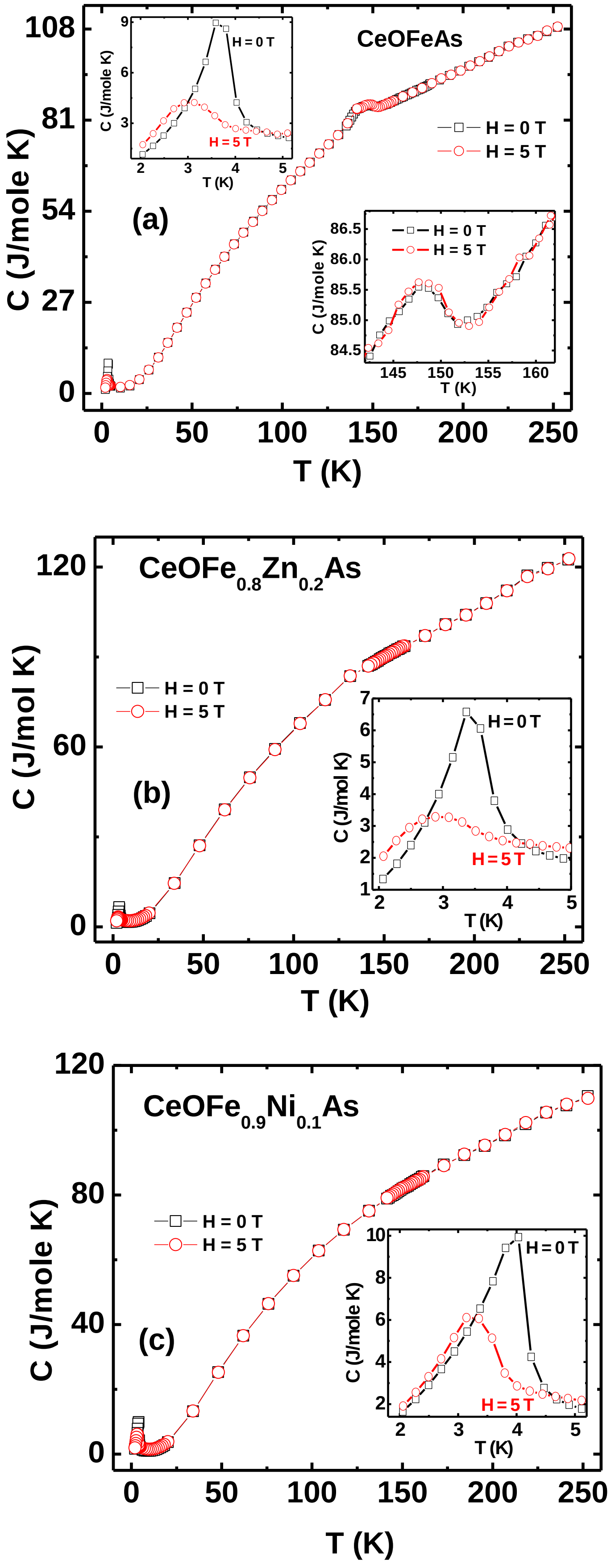

**Figure 4**

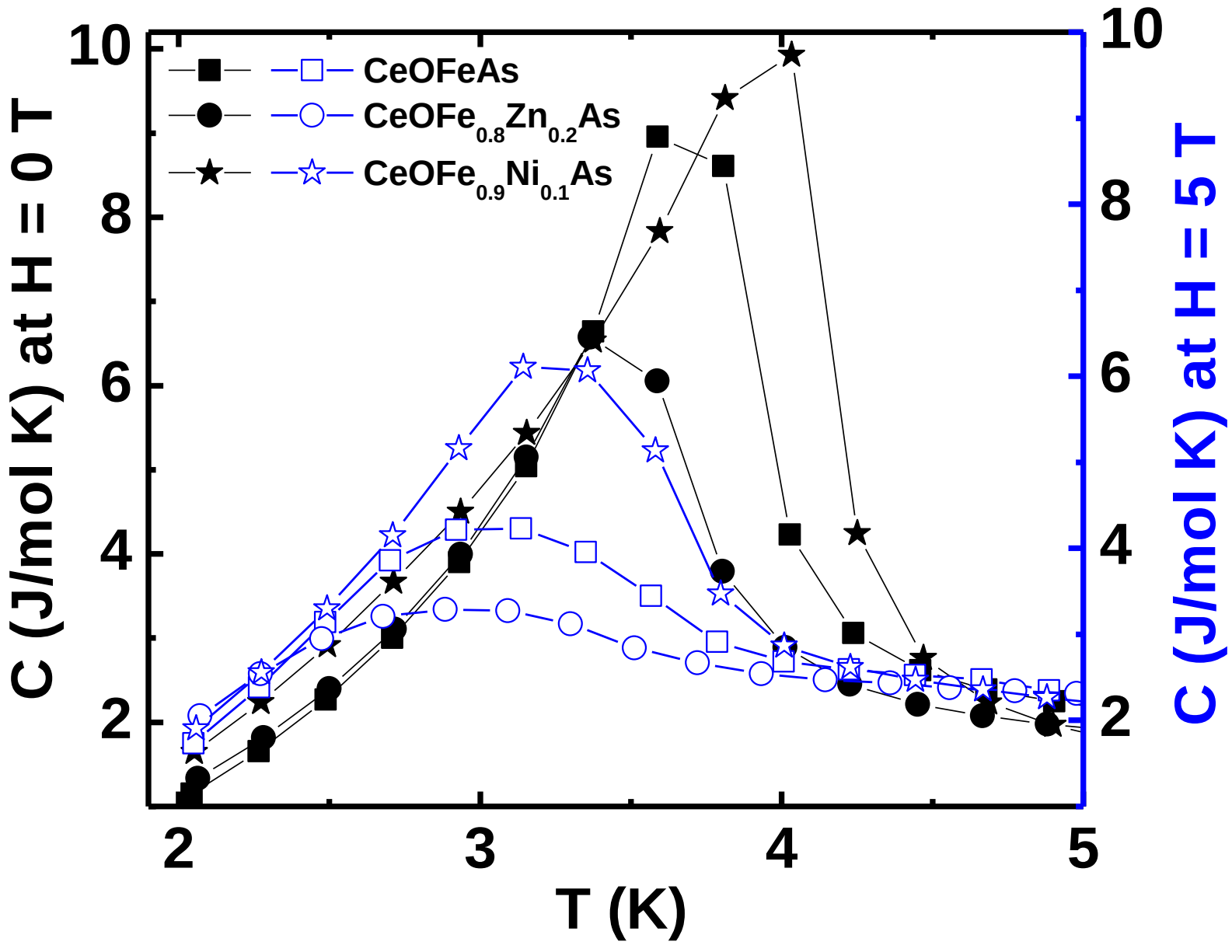

**Figure 5**

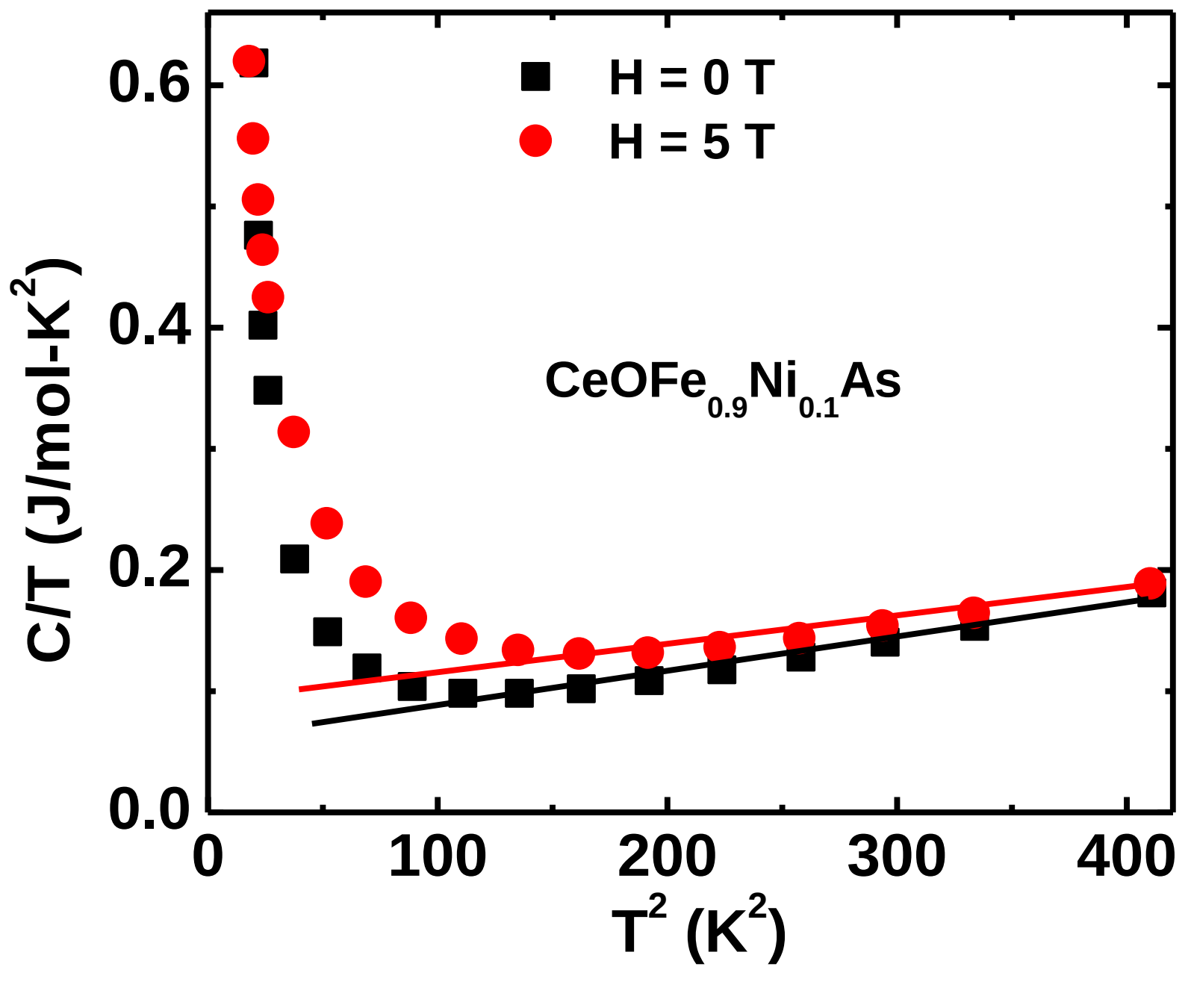

**Figure 6**

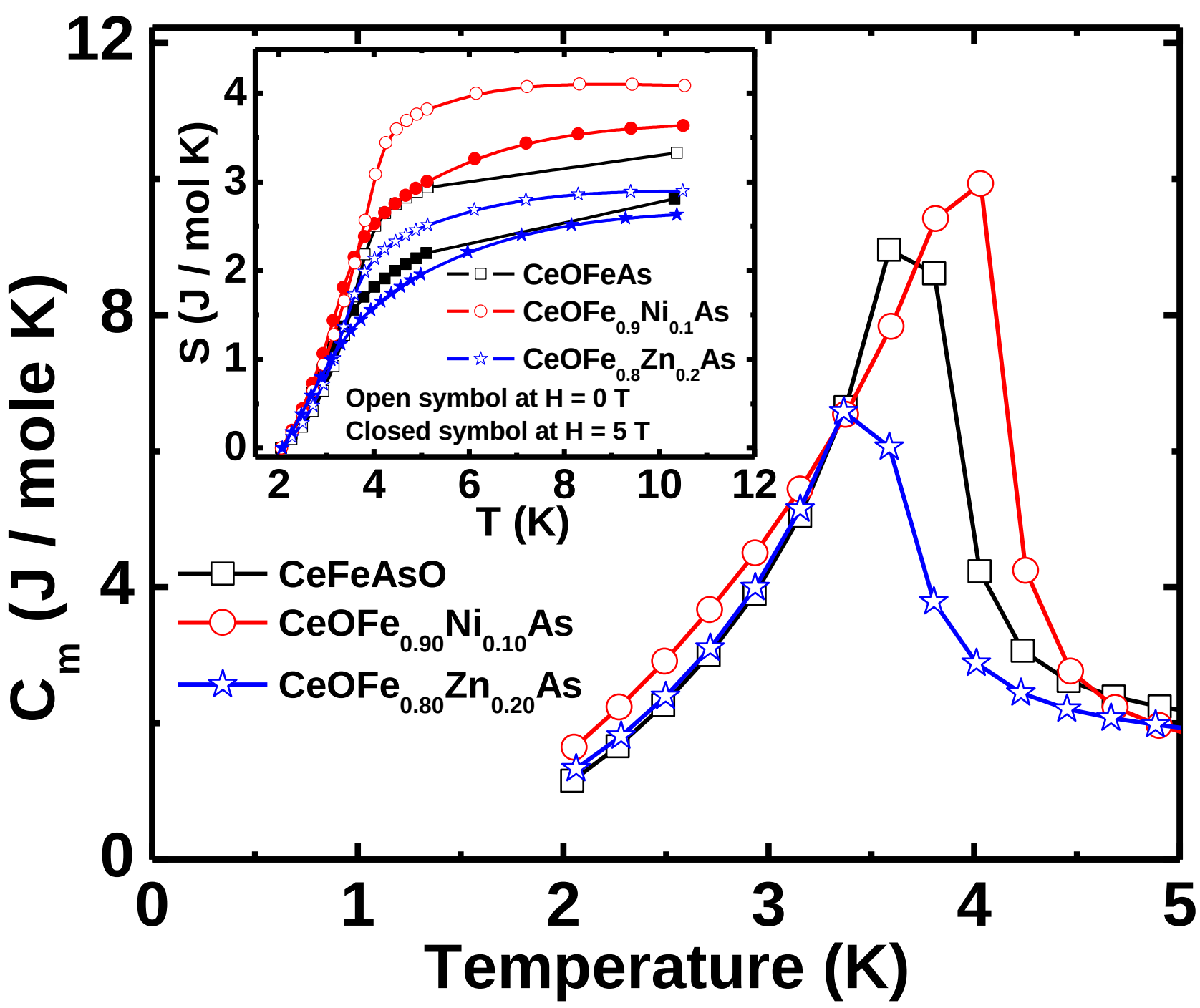